%% file: eprint.tex
\documentclass[12pt]{article}
\usepackage{graphicx}
\usepackage{amssymb}

\newcommand\pubnumber{}
\newcommand\pubdate{\today}

\def\israel{Racah Institute of Physics\\
The Hebrew University, Jerusalem, 91904, ISRAEL}
\def\japan{Department of Astronomy, Graduate School of Science\\
The University of Tokyo, 7-3-1, Hongo, 113-0033, JAPAN}
\def\email{\footnote{norita@astron.s.u-tokyo.ac.jp}}

\def\Title#1{\begin{center} {\Large #1 } \end{center}}
\def\Author#1{\begin{center}{ \sc #1} \end{center}}
\def\Address#1{\begin{center}{ \it #1} \end{center}}

\newcommand\pubblock{\rightline{\begin{tabular}{l} \pubnumber\\
         \pubdate  \end{tabular}}}
\newenvironment{Abstract}{\begin{quotation}  }{\end{quotation}}
\newenvironment{Presented}{\begin{quotation} \begin{center} 
             PRESENTED AT\end{center}\bigskip 
      \begin{center}\begin{large}}{\end{large}\end{center} \end{quotation}}
\def\Acknowledgements{\bigskip  \bigskip \begin{center} \begin{large}
             \bf ACKNOWLEDGEMENTS \end{large}\end{center}}
\input econfmacros.tex
\begin{document}
\begin{titlepage}
\pubblock

\vfill
\Title{Jet Luminosity from Neutrino-Dominated Accretion Flows in GRBs}
\vfill
\Author{Norita Kawanaka\email}
\Address{\israel}
\Address{\japan}
\vfill
\begin{Abstract}
A hyperaccretion disk around a stellar-mass black hole is a plausible model for the central engine that powers gamma-ray bursts (GRBs).  We estimate the luminosity of a jet driven by magnetohydrodynamic processes such as the Blandford-Znajek (BZ) mechanism as a function of mass accretion rate, the black hole mass, and other accretion parameters.  We show that the jet is most efficient when the accretion flow is cooled via optically-thin neutrino emission, and that its luminosity is much larger than the energy deposition rate through $\nu\bar{\nu}$ annihilation provided that the black hole is spinning rapidly enough.  Also, we find a significant jump in the jet luminosity at the transition mass accretion rate between the advection dominated accretion flow (ADAF) regime and the neutrino-dominated accretion flow (NDAF) regime.  This may cause the large variability observed in the prompt emission of GRBs.
\end{Abstract}
\vfill
\begin{Presented}
Huntsville Gamma-Ray Burst Symposium\\
Nashville, USA, April 14--18, 2013
\end{Presented}
\vfill
\end{titlepage}
\def\thefootnote{\fnsymbol{footnote}}
\setcounter{footnote}{0}

\section{Introduction}

The central engine of gamma-ray bursts (GRBs) that powers a relativistic jet is still unclear, but it is most likely to be a hyperaccreting black hole \cite{Narayan:1992iy, Narayan:2001qi}.  Such a system is expected to form after the gravitational collapse of a massive star or a merger of a neutron star binary \cite{Piran:1999kx}.  The accretion flow is extremely optically-thick with respect to photons and it cannot cool via radiation.  Instead, because of its high density and temperature, the flow cools via neutrino emission.  It is often called as a "neutrino-dominated accretion flow (NDAF)", and its structure and stability has been investigated by many authors \cite{Popham:1998ab, Narayan:2001qi, Matteo:2002ck, Kohri:2002kz, Kohri:2005tq, Gu:2006nu, Chen:2006rra, Kawanaka:2007sb, Liu:2007bca, Kawanaka:2011yb}.  There are two different processes that are most discussed in the literature: neutrino pair annihilation \cite{Eichler:1989ve} and magnetohydrodynamical mechanism such as Blandford-Znajek (BZ) process \cite{Blandford:1977ds}.  In this work we estimate the jet luminosity expected from a BZ mechanism in the context of a hyperaccretion flow, as well as its dependence on mass accretion rate, black hole mass, and other properties.  In the BZ process, the jet energy is extracted from a rotating black hole via magnetic field lines threading the black hole horizon.  In our work, we estimate that power using the maximal magnetic field sustainable on the horizon, which is assumed to be limited by the disk pressure near the innermost radius of the disk \cite{Krolik:2011fb, Krolik:2011sv}.  We also compare this luminosity with the energy deposition rate expected from neutrino pair annihilation above the accretion flow\footnote{For the detailed discussion, see \cite{Kawanaka:2012ub}.}.

\section{Model}

The Poynting luminosity expected from the BZ process is the order of $\sim c(B^2/8\pi)R_g^2$, where $B$ is the poloidal magnetic field strength on the horizon and $R_g=GM_{\rm BH}/c^2$ is the gravitational radius of the central black hole.  Following \cite{Krolik:2011fb}, we estimate the jet luminosity $L_{\rm jet}$ as
\begin{eqnarray}
L_{\rm jet}=f(a/M_{\rm BH})c(B^2/8\pi)R_g^2,
\end{eqnarray}
where $f(a/M_{\rm BH})$ is an increasing function of $\left|a/M_{\rm BH}\right|$ whose exact form depends on the geometry of the magnetic field.  According to \cite{Beckwith:2007sr}, low-order poloidal topologies of magnetic field are the most favorable to strong jets.  When the topology is optimal and $a/M_{\rm BH} \gtrsim 0.9$, the $f(a/M_{\rm BH})$ arising from a dynamically self-consistent field structure can be $\gtrsim 0.05$, and even as large as unity for spin parameters closer to unity \cite{Hawley:2005xs}.  In the following discussion we assume that $f(a/M_{\rm BH})$ is the order of unity.

In estimating the magnetic field strength on the horizon, we follow Beckwith et al. \cite{Beckwith:2009rm} who argued that the magnetic pressure near the horizon may be limited by the inner disk pressure:
\begin{eqnarray}
\frac{B^2}{8\pi}\approx p_{\rm disk}(R_{\rm in}),
\end{eqnarray}
where $R_{\rm in}$ is the radius of the pressure reference point.

To evaluate the midplane pressure of the hyperaccretion flow, we solve the disk structure in the innermost region ($R\sim O(R_g)$), where the jet is expected to be launched.  We assume that both the disk dynamics and the gravitational potential are Newtonian, and that the system is stationary and axisymmetric.  We adopt the Shakura-Sunyaev formalism for the disk structure\cite{Shakura:1972te}, which formally applies only to thin disks.  When the disk is advection-dominated (see below) and becomes geometrically thick, we use a dimensional analysis in our analytic estimates.

Here, we present the basic disk equations for the density $\rho$, the temperature $T$, and the scale height $H$.  These are the expressions for mass conservation, angular momentum conservation, energy conservation, and hydrostatic balance applied at $R_{\rm in}$:
\begin{eqnarray}
\dot{M}&=&-2\pi R_{\rm in}\Sigma v_R, \\
2\alpha H p_{\rm disk}&=&\frac{\dot{M}\Omega(R_{\rm in})}{2\pi}, \\
Q^+&=&Q^-, \\
\frac{p_{\rm disk}}{\rho}&=&\Omega(R_{\rm in})^2 H^2,
\end{eqnarray}
where $\Sigma$, $v_R$, $\Omega(R)$ and $\alpha$ denote the surface density ($=2\rho H$), radial velocity, angular velocity ($=GM_{\rm BH}/R^3$), and ratio of integrated stress to integrated pressure, respectively.  Here $Q^+$ is the heating rate per unit area, and $Q^-$ is the cooling rate per unit area, including both neutrino and advective cooling: $Q^-=Q_{\nu}^-+Q_{\rm adv}^-$.  For the neutrino cooling, we take into account electron/positron capture on nucleons, electron-positron pair annihilation, and other various neutrino emission processes (see \cite{Kawanaka:2012ub}).

By using the analytic approximation for the neutrino cooling and the disk pressure, we can distinguish different regimes of the solution for the disk structure.  

1. When the mass accretion rate is not large enough for the disk to cool efficiently via neutrino emission, the accretion flow is advection-dominated and the pressure is dominated by radiation (i.e. photons and relativistic pairs).  In this regime the jet luminosity is proportional to mass accretion rate.  The accretion flow is advection-dominated as long as the neutrino emissivity per unit volume is smaller than the advective cooling rate.

2.  If the mass accretion rate is larger than $\gtrsim 0.01M_{\odot}{\rm s}^{-1}$, the disk is mainly cooled by neutrino emission via electron/positron capture onto nucleons, and the pressure is dominated by baryonic gas (see \cite{Kawanaka:2007sb}).  In this regime, as in the previous regime, the jet luminosity is proportional to mass accretion rate.  However, the normalization is a few times larger than in the previous regime.  This means that the jet luminosity has a step-function-like behavior at the transition mass accretion rate between these two regimes.

3. If the mass accretion rate is sufficiently large ($\gtrsim 0.04M_{\odot}{\rm s}^{-1}$), the accretion flow becomes optically-thick with respect to neutrinos, and neutrino cooling should be described with diffusion approximation.  In this regime, the jet luminosity is proportional to $\dot{M}^{2/3}$.

4. If the mass accretion rate is very large ($\gtrsim 4M_{\odot}{\rm s}^{-1}$), neutrinos are completely trapped in the accretion flow.  In this regime, the accretion flow is advection-dominated, and the pressure is dominated by radiation and neutrinos.  As in the first regime, the jet luminosity is proportional to mass accretion rate.

\section{Results and Discussion}
Fig. 1 depicts the jet luminosities expected from the BZ mechanism and neutrino pair annihilation as functions of mass accretion rate.  Obviously, the BZ mechanism is much more efficient than neutrino pair annihilation in energizing a jet.  In addition, the jet luminosity based on the BZ scenario has a significant discontinuity at the transition between the first and second regime.  Fig. 2 depicts the jet luminosities with various parameter sets $(M_{\rm BH}, \alpha, R_{\rm in})$, we can also see a similar discontinuity in each curve.   The jet luminosity around this jump is $\sim 10^{50-51}{\rm erg}~{\rm s}^{-1}$, and this value is just similar to that inferred from observed GRBs.  In addition, the drop in the jet luminosity (a factor of $\sim 3-5$) may lead to the variability observed in the prompt emissions, whose amplitude is a factor of a few, or the steep decay in the X-ray afterglows of GRBs.

\begin{figure}[tbp]
\centering
\includegraphics[height=1.5in]{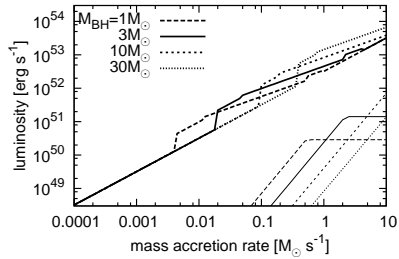}
\caption{Jet luminosities expected from the BZ mechanism (thick lines) and neutrino pair annihilation (thin lines) as functions of mass accretion rate for several different radii.}
\label{fig:1}
\end{figure}

Strictly speaking, the BZ jet luminosity is larger than the energy deposition rate via neutrino pair annihilation only when $f(a/M_{\rm BH})\gtrsim 0.01$, which means that the black hole spin is large and the configuration of the magnetic field is relevant.  However, in the optimal range of mass accretion rates, this bound is relaxed another order of magnitude and that makes the necessary conditions for black hole spin and magnetic field far from extreme.

\begin{figure}[tbp]
\centering
\includegraphics[height=3in]{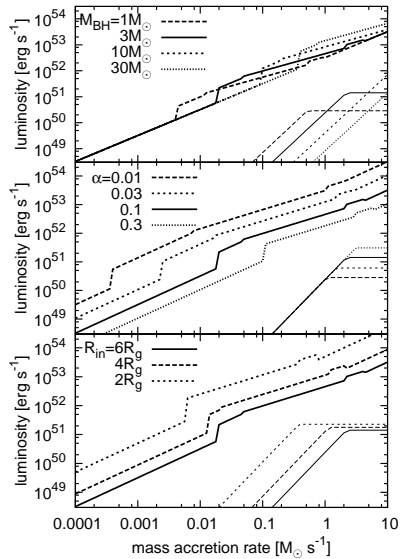}
\caption{Jet luminosities as functions of mass accretion rate.  Their dependencies on the black hole mass, $\alpha$, and the inner radius of an accretion flows are shown.}
\label{fig:2}
\end{figure}

\Acknowledgements
This work is supported by an Advance ERC grant, and the numerical calculations were carried out on SR16000 at Yukawa Institute for Theoretical Physics at Kyoto University.

\end{document}

%% file: econfmacros.tex
\def\beq{\begin{equation}}
\def\eeq#1{\label{#1}\end{equation}}
\def\eeqn{\end{equation}}

\def\beqa{\begin{eqnarray}}
\def\eeqa#1{\label{#1}\end{eqnarray}}
\def\eeqan{\end{eqnarray}}

\let\bar=\overbar

\def\Dslash{\not{\hbox{\kern-4pt $D$}}}
\def\dslash{\not{\hbox{\kern-2pt $\del$}}}

\def\msb{{\bar{\ssstyle M \kern -1pt S}}}